\documentclass{ws-procs10x7}
\usepackage{balance}
\newcommand{\ET}{{\rm E}_{\scriptscriptstyle\rm T}}
\newcommand{\MET}{\mbox{$\raisebox{.3ex}{$\not$}\ET$}}

\newcolumntype{d}[1]{D{.}{.}{#1}}

\makeindex
\begin{document}

\title{Search for Low Mass Higgs at the Tevatron}

\author{Benjamin Kilminster on behalf of the CDF and D\O \ collaborations}

\address{Department of Physics \& Astronomy, Ohio State University, Columbus,
OH 43210, USA\\$^*$E-mail: bjk at fnal.gov}

\twocolumn[\maketitle\abstract{We present CDF and D\O \ searches for a Standard
Model Higgs boson produced associatively with a $W$ or $Z$ boson at
$\sqrt{s}=1.96~$ TeV using up to 1 fb$^{-1}$ of analyzed Tevatron data
collected from February 2002 to February 2006.  For Higgs masses less than 135 GeV/c$^2$, as is
favored by experimental and theoretical constraints, $W^{\pm}H \to l^{\pm}\nu
b\bar{b}$, $ZH \to l^+l^-b\bar{b}$, and $ZH \to \nu\bar{\nu} b\bar{b}$ are
the most sensitive decay channels to search for the Higgs boson. Both CDF and D\O \ have
analyzed these three channels and found no evidence for Higgs production, and
therefore set upper limits on the Higgs production cross-section. While the
analyses are not yet sensitive to Standard Model Higgs production,
improvements in analysis techniques are increasing sensitivity to the Higgs
much faster than added luminosity alone.}
]

\section{Introduction}
\label{sec:intro}

The Higgs boson is the only particle predicted by the Standard Model which
has not yet been detected.  It holds a special place in the Standard Model as
the only scalar boson and as the particle which gives mass to the fermions,
its coupling strength increasing as a function of their mass.  Its predicted
couplings to other Standard Model particles has allowed its mass to be
indirectly constrained by precision electroweak data, most importantly the
top quark mass and the $W$ mass.  The new combined Tevatron top quark mass of 171.4
$\pm$ 2.1 GeV/c$^2$ \cite{top} and the new combined $W$ boson mass of 80.376 $\pm$ 0.033
GeV/c$^2$ \cite{W} have indirectly constrained m$_H$ to 85 +$^{39}$ -$_{28}$
GeV/c$^2$, yielding a 95\% CL upper limit of 166 GeV/c$^2$ \cite{lephiggs}.
Direct searches for the Higgs at LEP have excluded the Higgs up to 114.4 GeV/c$^2$
\cite{Barate:2003sz}.  There is then clear motivation for searching for a low
mass Higgs just above 114.4 GeV/c$^2$.

CDF and D\O \ at the Tevatron are currently the only experiments capable of
searching for the Higgs boson. Above 135 GeV/c$^2$, $H \to W^+W-$ is the
dominant Higgs decay mode, while below 135 GeV/c$^2$, the focus for this
proceeding, Higgs decays predominantly $H \to b\bar{b}$.  The dominant mode
of production at the Tevatron is directly via $p\bar{p} \to H$ ($\sigma = $ 0.8 pb) which is about
four and seven times larger than the associative modes of $p\bar{p}\to W^*
\to WH$ and $p\bar{p} \to Z* \to ZH$ for M$_H = $ 115 GeV/c$^2$.  The direct production
signature of $H \to b\bar{b}$ is overwhelmed by QCD dijet backgrounds. The
associative modes are most sensitive for discovering a low mass Higgs since
the background rate is much smaller due to the reconstruction of the $Z$ or $W$. 

CDF and D\O \ have searched for the low mass Higgs in three associative
channels, $W^{\pm}H \to l^{\pm}\nu b\bar{b}$, $ZH \to l^+l^-b\bar{b}$, and $ZH
\to \nu\bar{\nu} b\bar{b}$, where $l$ can be an electron or a muon. 

\section{Identifying $b$ quarks from Higgs decay}
\label{sec:bquarks}

Each of the three low mass searches share the decay $H \to b\bar{b}$ which
has a branching ratio of 73\% for m$_H = $ 115 GeV/c$^2$.  Therefore,
$b$-quark identification is a key element in event selection.  Jets
originating from $b$ quarks can be ``$b$-tagged'' by identifying the $b$
decay vertex, which is displaced from the collision vertex due to the large
$b$ lifetime.  At CDF, the $b$-tag is 42 \% efficient for jets within $\eta
<$ 1.0 and with $E_T >$ 15 GeV.  The ``mistag'' rate for light-quark jets to
be $b$-tagged is about 1\%.  Since the probability of two mistags is small,
and $H \to b\bar{b}$ has two possible $b$-tags, both CDF and D\O \ divide
events into two separate samples by the number of $b$-tags for maximal signal to background
sensitivity.  At CDF a Neural Network is used in the $WH$ analysis to further
divide $b$-tagged jets into those most likely coming from $b$'s, $c$'s, and
mistags.

\begin{figure}[b]
\centerline{\psfig{file=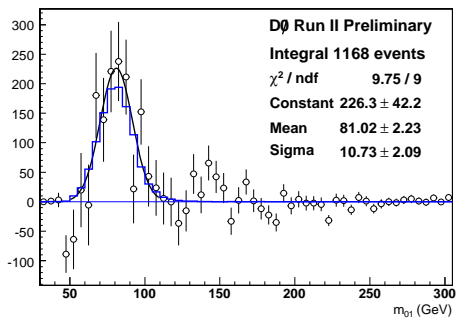,width=2.7in}}
\caption{ The extracted $Z\to b\bar{b}$ mass peak in single $b$-tagged events from D\O \ .}
\label{fig:zbb}
\end{figure}

A benchmark for Higgs searches is reconstructing $b\bar{b}$ resonances from
$b$-tagged events.  D\O \ finds evidence for a $Z \to b\bar{b}$ signal of 1168
events from a QCD background of almost a million events, measuring M$_Z$ =
81.0 $\pm$ 2.2 GeV/$c^2$, compared to an expected measured value from pseudo-data
of 83 $\pm$ 2 GeV/$c^2$ (Figure \ref{fig:zbb}).

\section{Higgs Search in $W^{\pm}H \to l^{\pm}\nu b\bar{b}$}
\label{sec:wh}

\begin{figure}[b]
\centerline{\psfig{file=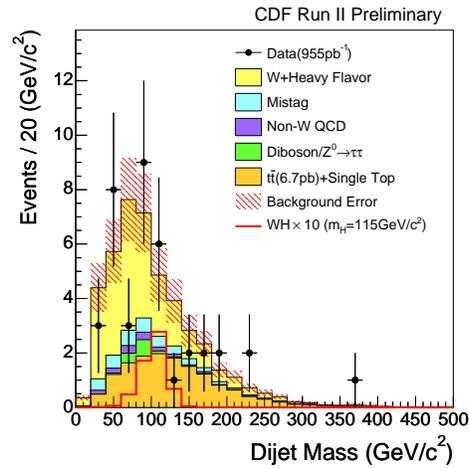,width=2.6in}}
\caption{The CDF analysis of $WH \to l\nu b\bar{b}$ showing the dijet mass distribution used to search for the higgs resonance.}
\label{fig:whjj}
\end{figure}

\begin{figure}[b]
\centerline{\psfig{file=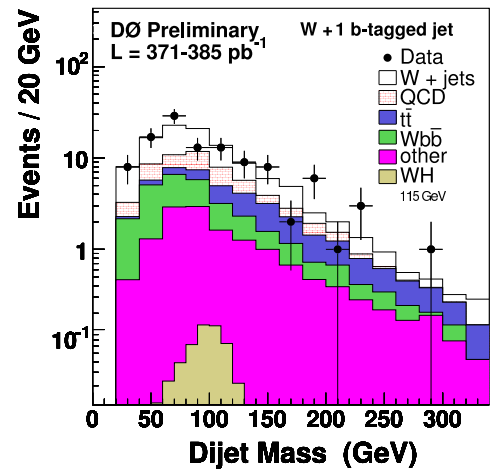,width=2.6in}}
\caption{The D\O \ analysis of $WH \to l\nu b\bar{b}$ showing the dijet mass distribution used to search for the higgs resonance.}
\label{fig:d0whjj}
\end{figure}

The signature from $W^{\pm}H \to l^{\pm}\nu b\bar{b}$ of a high transverse
momentum (P$_T$) lepton, missing transverse energy ($\MET$) from the
neutrino, two or more high transverse energy ($\ET$) jets, and one or more
$b$-tags, is well understood since it is almost the same channel in which CDF
and D\O \ measure $t\bar{t}$ production.  The main difficulties for the $WH$
analysis are estimating the dominant $W$ plus heavy flavor jets background which
can be seen in the dijet mass distributions that are used to search for the
Higgs resonance in Figures
\ref{fig:whjj} and \ref{fig:d0whjj}.

\section{Higgs Search in $ZH \to \nu\bar{\nu} b\bar{b}$}
\label{sec:zhnunubb}

\begin{figure}[b]
\centerline{\psfig{file=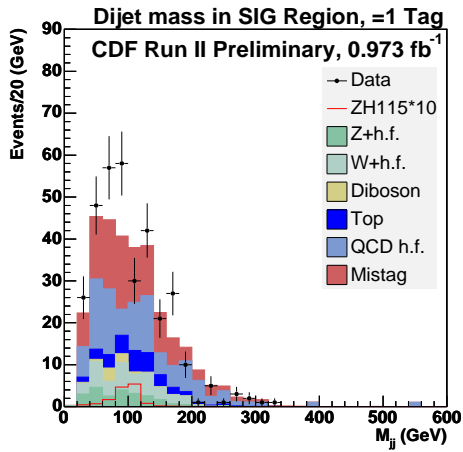,width=2.6in}}
\caption{ The CDF analysis of $ZH \to \nu\bar{\nu} b\bar{b}$ showing the dijet mass distribution used to search for the higgs resonance.}
\label{fig:cdfzhnunubb}
\end{figure}

\begin{figure}[b]
\centerline{\psfig{file=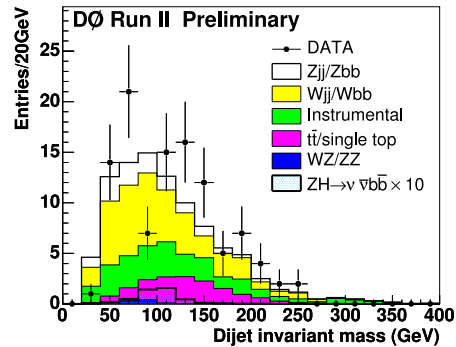,width=2.6in}}
\caption{ The D\O \ analysis of $ZH \to \nu\bar{\nu} b\bar{b}$ showing the dijet
mass distribution used to search for the higgs resonance.}
\label{fig:d0zhnunubb}
\end{figure}

The $ZH \to \nu\bar{\nu} b\bar{b}$ signature of two high $E_T$ jets with at
least one $b$-tag recoiling against $\MET$ from Z decays to neutrinos
is the most sensitive signature to low mass Higgs production at the Tevatron.
At CDF, the search in the $\MET$ + jets channel includes $WH$ events which
did not pass lepton identification requirements.  At D\O \ , a separate
search for $WH$ in the $\MET$ + jets channel is made.  The largest
backgrounds are due to jet energy mismeasurements of heavy flavored dijet QCD
production, as well as $W$+jets and $Z$+jets where either the neutrinos or
leptons are not reconstructed and yield $\MET$.  Checks are made to verify
the modeling of these backgrounds in two control regions sensitive to these
QCD and electroweak backgrounds.  The dijet mass distributions in the signal
region for CDF and D\O \ are shown in Figures
\ref{fig:cdfzhnunubb} and
\ref{fig:d0zhnunubb}.




\section{Higgs Search in $ZH \to l^+l^-b\bar{b}$}
\label{sec:zhllbb}

\begin{figure}[b]
\centerline{\psfig{file=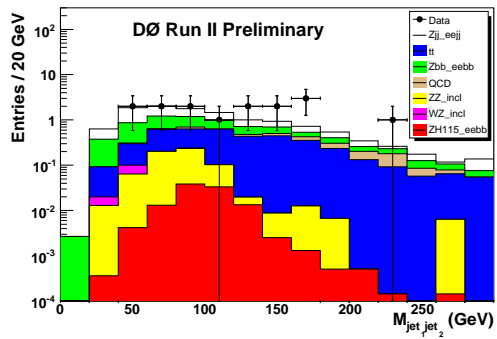,width=2.6in}}
\caption{ The D\O \ analysis of $ZH \to l^+l^-b\bar{b}$ ($ee$) showing the dijet
mass distribution used to search for the higgs resonance. }
\label{fig:d0zhllbb}
\end{figure}

\begin{figure}[b]
\centerline{\psfig{file=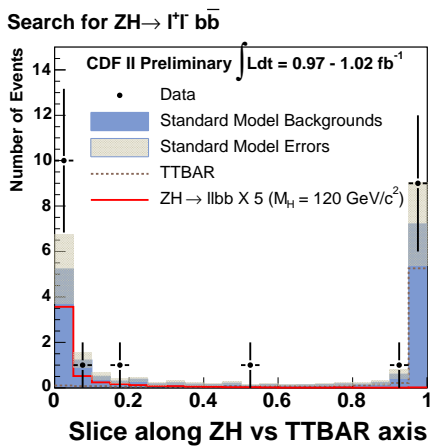,width=2.5in}}
\caption{ The CDF analysis of $ZH \to l^+l^-b\bar{b}$ ($ee$,$\mu\mu$) showing
a slice of the 2-D Neural Network discriminant used to search for the higgs
resonance along the $ZH$ vs $t\bar{t}$ axis. }
\label{fig:cdfzhllbb}
\end{figure}

The $ZH \to l^+l^-b\bar{b}$ channel has the smallest event yield, but
requiring two identified leptons to reconstruct the $Z$ mass results in the
cleanest signal.  The D\O \ analysis uses the dijet mass to discriminate signal
from the backgrounds dominated by $Z+$ jets (Figure \ref{fig:d0zhllbb}). The CDF analysis uses a two
dimensional Neural Network discriminant based on nine kinematic variables in
order to maximize signal discrimination from $Z+$ jets and subdominant $t\bar{t}$ background (Figure \ref{fig:cdfzhllbb}).

\section{Summary of Search Results}
\label{sec:sum}

No significant signal excess was measured in any of the CDF and D\O \
channels and therefore upper limits are set on Higgs production.  These
limits are summarized in Table \ref{tab:summary}.  Note that CDF
combines its $ZH$ and $WH$ signals in the $\MET$ + jets channel, while D\O
\ has separate $\MET$ + jets results for each signal.  Also, D\O \ reports
$Z \to ee $ and $Z \to \mu\mu$ as separate analyses while CDF does a combined
fit in both channels.

These results are combined according to their statistical and systematic
correlations into an upper limit on Higgs production as a function of mass.
At 115 GeV/c$^2$, the observed (expected) upper limit is 10.4 (7.6) times larger than the Standard
Model prediction\cite{tevhiggs}.  With the full D\O
\ 1 fb$^{-1}$ dataset, these expected limits would be 25\% better.

\section{Conclusions}
\label{sec:conc}

The prospects for Higgs discovery or exclusion at the Tevatron depend on
large integrated luminosity, advanced techniques for extracting Higgs
candidates from background, and the timing of the transition to quality
physics results at the LHC.

The Tevatron is well on its way to delivering the design goal of 8 fb$^{-1}$,
after delivering more than 1.6 fb$^{-1}$ of data, and recently surpassing the
integrated luminosity rate necessary for delivering 4 fb$^{-1}$.

In this proceeding, we reported for the first time the addition of the CDF
and D\O \ $ZH \to l^+l^-b\bar{b}$ channels to the combined Higgs search.  We
also report for the first time the updated 1 fb$^{-1}$ results for CDF's
$W^{\pm}H \to l^{\pm}\nu b\bar{b}$, and $ZH \to \nu\bar{\nu} b\bar{b}$.

With 1 fb$^{-1}$ of data, the Tevatron is about a factor of 6 away from the
Standard Model in terms of expected limits for a low mass Higgs.  As new data
are accumulating, Higgs analysis techniques are also becoming more advanced, making use
of Neural Networks for $b$-tagging as well as kinematic separation of signal
from background.


\begin{table}
\tbl{Summary of CDF and D\O \ Higgs upper limits for $M_H =$ 115 GeV/c$^2$. L/SM is the measured upper limit of $\sigma_{ZH}$ or $\sigma_{WH}$ divided by its Standard Model NLO Cross-Section. }
{\begin{tabular}{@{}lccc@{}}
\toprule
Expt & Channel   & ${\cal L}$ (fb$^{-1}$)  & L/SM \\ \colrule
CDF  & $WH \to l\nu b\bar{b}$ & 1   & 25 \\
D\O \   & $WH \to l\nu b\bar{b}$ & 0.38 & 18 \\
CDF  & $Z/W H \to \MET b\bar{b}$    & 1    & 14 \\
D\O \   & $ZH \to \MET b\bar{b}$    & 0.26 & 41 \\
D\O \   & $WH \to \MET b\bar{b}$    & 0.26 & 55 \\
CDF  & $ZH \to llb\bar{b}$           & 1    & 28 \\
D\O \   & $ZH \to eeb\bar{b}$           & 0.39 & 100 \\
D\O \   & $ZH \to \mu\mu b\bar{b}$           & 0.32 & 139 \\
\\
Comb & All & 0.3-1.0  & 10.4 \\
\botrule
\end{tabular}}
\label{tab:summary}
\end{table}

\end{document}